# Ending AIDS in Gabon: How long will it take? How much will it cost?

Brian G. Williams,† Eleanor Gouws* and David Ginsburg¶

† South African Centre for Epidemiological Modelling and Analysis (SACEMA), Stellenbosch, South Africa
* UNIADS Regional Office for Eastern and Southern African, Johannesburg, South Africa

Correspondence to BrianGerardWilliams@gmail.com

**Abstract**

The prevalence of HIV in West Africa is lower than elsewhere in Africa but Gabon has one of the highest rates of HIV in that region. Gabon has a small population and a high *per capita* gross domestic product making it an ideal place to carry out a programme of early treatment for HIV. The effectiveness, availability and affordability of triple combination therapy make it possible to contemplate ending AIDS deaths and HIV transmission in the short term and HIV prevalence in the long term. Here we consider what would have happened in Gabon without the development of potent anti-retroviral therapy (ART), the impact that the current roll-out of ART has had on HIV, and what might be possible if early treatment with ART becomes available to all.

We fit a dynamic transmission model to trends in the adult prevalence of HIV and infer trends in incidence, mortality and the impact of ART. The availability of ART has reduced the prevalence of HIV among adults not on ART from 4.2% to 2.9%, annual incidence from 0.43% to 0.27%, and the proportion of adults dying from AIDS illnesses each year from 0.36% to 0.13% saving the lives of 2.3 thousand people in 2013 alone. The provision of ART has been highly cost effective saving the country at least $18 million up to 2013.

## Introduction

The prevalence of HIV infection in Gabon has increased rapidly to a peak in the year 2000 and has fallen slightly since then. Gabon now has a reasonable proportion of people on ART and this has undoubtedly saved many lives and significantly reduced transmission. If the Gabonese Government follows the current recommendations of the World Health Organization (WHO), the International AIDS Society (IAS) and the Department of Health and Human Services (DHHS) this has the potential to end the epidemic. If this is to be done it will be important to estimate the impact that this will have on the epidemic and how much it will cost. Here we use UNAIDS estimates of the trend in the prevalence of HIV and the number of people currently receiving ART in Gabon. We fit these data to a dynamical model to estimate current, and to project future, trends in prevalence, incidence, treatment needs and deaths. We estimate the cost of HIV/AIDS to the country including the cost of providing drugs and providing support to people on ART, the cost of hospitalization and access to primary health care facilities. We do not include the cost to the county of deaths among young adults so that these calculations are likely to be conservative with respect to costs.

## Data

The model is fitted to data provided by UNAIDS on the trend in the prevalence of HIV. Costs are based on a recent study for South Africa including the cost of hospitalization, primary health care and treatment.[1] Data in this regard for Gabon would provide better estimates. The costs are given in Table 1. All costs are in 2013 US$ discounted at 3% per year.

## Model

The model is a standard dynamical model discussed in detail elsewhere.[4,5] In brief, the model includes uninfected people who are susceptible to infection while infected people go through four stages of infection to death as this gives an accurate match to the known Weibull survival for people infected with HIV but not ART.[6] To account for heterogeneity in the risk of infection the transmission parameter declines exponentially with the prevalence of HIV.[4,5]

Table 1. Cost of days in hospital, primary health care visits and counselling and testing from Granich *et al.*[1] Drug costs are current prices for first line drugs in South Africa.[3] The cost of care and support covers community based care, the cost of a death is set to the GNI for Gabon.[2,3] Costs are for the year 2013; future costs are discounted at 3.5% per year.

| Item | Cost ($ *p.a.*) |
| --- | --- |
| Inpatient days not on ART | 568 |
| Inpatient days on ART | 138 |
| Primary health care visits not on ART | 154 |
| Primary health care visits on ART | 269 |
| Counselling and testing per test | 11 |
| Cost of the death of a young adult | 8,000 |
| Drug costs | 100 |
| Community support and care | 150 |

## Fitting the model

We first fit the model to the prevalence data without including ART to get a null model against which to compare the impact of ART. We vary the prevalence of infection in 1980, which determines the timing of the epidemic, the rate of increase and the rate at which the risk of infection declines as the prevalence increases as this determines the peak value of the prevalence. This gives the parameter values in Table 2.

To model the current level of ART provision we assume that certain proportions of people in the fourth stages of HIV infection are started on ART. We assume that coverage increases logistically and vary the rate and timing of the increase and the proportion of people starting treatment in each stage to match the reported coverage of ART. To explore the impact of active case finding we again assume that coverage increases logistically at a realistic rate, timing and asymptotic coverage. We assume that all those that are found to be HIV-positive, in any stage of infection, are eligible for treatment. The parameter values are given in Table 3.

Table 2. The best fit parameter values for the prevalence of HIV. The birth and transition rates are fixed; the other parameters are varied to optimize the fit.

| Parameter | Value |
| --- | --- |
| Birth rate/yr | 0.029 |
| Background mortality/yr | 0.018 |
| Force of infection/yr | 0.375 |
| Prevalence at which transmission is halved | 0.034 |
| Transition rate/yr between states off ART | 0.348 |
| Transition rate/yr between ART states on ART | 0.087 |

Table 3. The model parameters are for logistic functions that determine coverage and testing rates. Coverage gives the asymptotic value, rate is the exponential rate of increase, 'half-max' gives the year when coverage reaches half the maximum value. Passive case finding applies to those that present in late stages of HIV. Active case finding applies to all infected people.

| | Parameter | Value |
| --- | --- | --- |
| Behaviour change | Reduction | 0.38 |
| | Rate *per annum* | 0.93 |
| | Half-max. (year) | 2002.06 |
| Passive case Finding: Stage 4 | Coverage | 0.50 |
| | Rate *per annum* | 0.90 |
| | Half-max. (year) | 2004.70 |
| Active case finding: testing | Coverage | 0.90 |
| | Rate *per annum* | 2.00 |
| | Half-max. (year) | 2014.0 |
| | Test interval (yrs) | 1.00 |
| Active case finding: take-up | Coverage | 0.90 |
| | Rate *per annum* | 2.00 |
| | Half-max. (year) | 2014.00 |

## Results

The fitted values of the parameters are given in Table 2 and Table 3 and the fitted data and implied trends are shown in Figure 1. The graphs in Figure 1 give, from left to right, the prevalence of infection, the annual incidence of infection, of treatment and the mortality, and the costs, and from top to bottom the counterfactual that would have happened without ART, the impact of the current level of ART, and the predicted impact of universal access to early treatment.

## Without ART

Figure 1A gives the prevalence, Figure 1B the implied incidence (red line) and mortality (black line) and Figure 1C the implied costs of letting people die without ART (blue line) and the cost of in- and out-patient care[1] (black line). The prevalence rises rapidly to a steady state, falls slightly, and then remains fairly constant after 2002 (Figure 1A). In order to fit the drop in the observed prevalence we have to assume a degree of behaviour change leading to a 38% decline in risk, as indicated in Table 3. The incidence peaks, declines as people reduce their risky behaviour and then levels off. The mortality rises about ten years after the incidence reflecting the mean life-expectancy of people with HIV (Figure 1B).

Without ART about 0.4 % of adults, or about 50k people, would have died in the years up to 2103 from AIDS and AIDS related illnesses (Figure 2B). The annual cost to society of providing clinical care and support for people with AIDS related illnesses would have peaked at $12 per adult in 2006, have fallen to about $9 per adult in 2021 (Figure 1C) and then continued to rise because of the discounting rate for future costs. The cumulative cost of care and support for those infected with HIV would have reached US$191M in 2013 and US$355M by 2023.

## Progress so far

The provision of ART on a significant scale in the public sector in South Africa expanded rapidly. Allowing for the provision of ART, Figure 1D gives the prevalence of those not on ART (red line), on ART (green line) and the total (blue line); Figure 1E gives the incidence of HIV (red line), the rate at which people start ART (green line), the mortality of those on ART (black line) and of those not on ART (grey line). Figure 1F gives the mortality of people not on ART (red line), on ART (green line) and all deaths (blue line). In Figure 1F the black line includes the cost of in- and out-patient care.[1] Currently, about one-quarter of all HIV positive adults are currently believed to be on ART. While about 4.2% of adults are currently infected with HIV (Figure 1D: blue line), the prevalence among adults not on ART (Figure 1D red line) has fallen to about 2.9% of adults with the remaining 1.3% on ART and if the current policy were to continue the prevalence in adults infected with HIV but not on treatment would fall to about 2%.

As expected there has been a similarly large impact on incidence which has fallen from a peak of 0.6% *per annum* to 0.3% *per annum* (Figure 1E: red line) with the rate at which people start ART peaking at 0.4% *per annum* in 2007 (Figure 1E: green line) but falling rapidly after that. AIDS related deaths among adults peaked in 2006 at about 0.3% *per annum* but by 2013 has fallen to 0.1% *per annum* (Figure 1E: black line). Without ART about 3.2 thousand people would have died in 2013; the provision of ART has reduced this to 2.1 thousand saving more then one thousand lives in 2013 alone. People will, of course, continue to die on ART (Figure 1E: grey line) but from natural causes other than HIV.

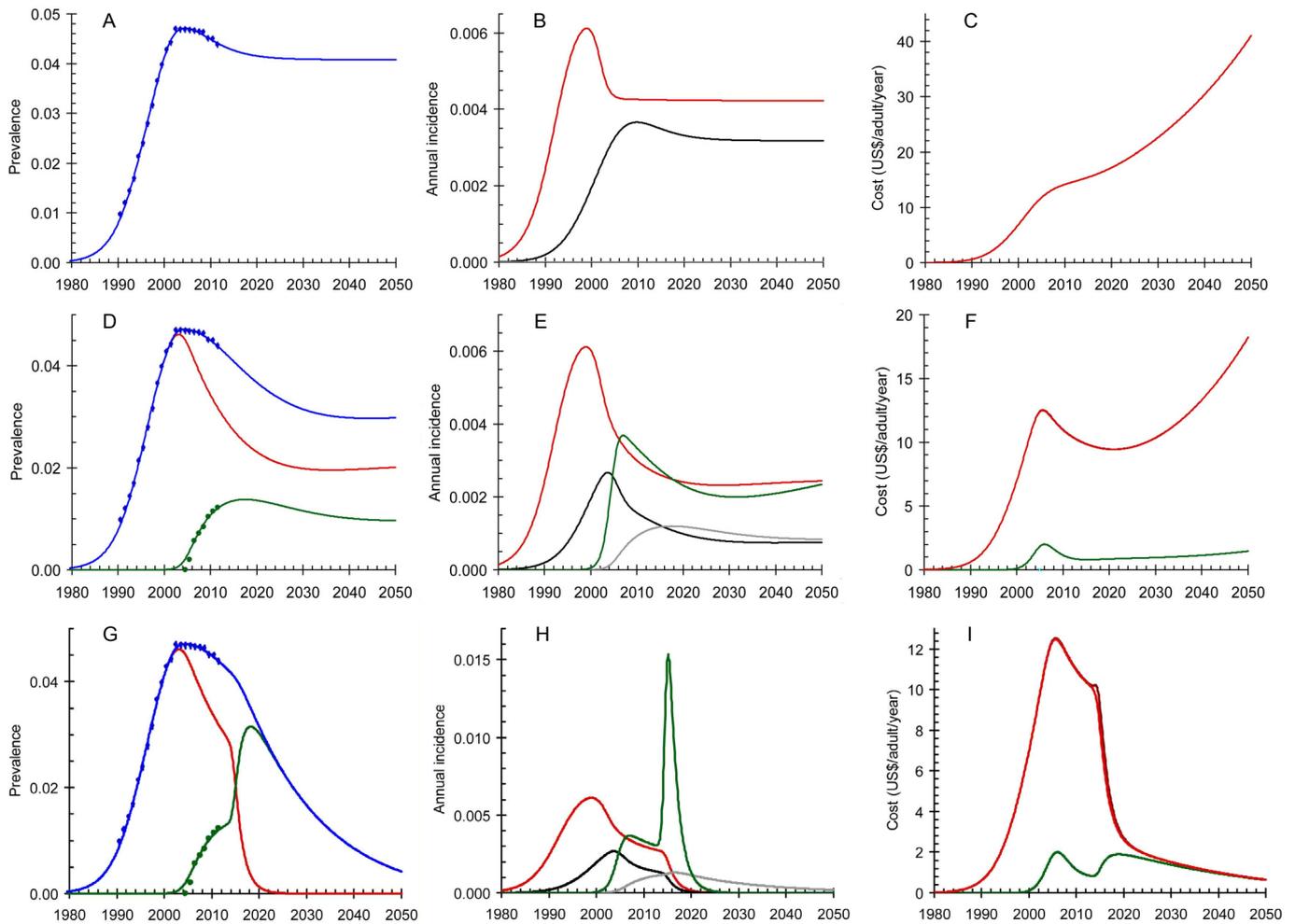

Figure 1. *Top:* baseline scenario without ART. *Middle:* current level of treatment. *Bottom:* early treatment (see text for details). *Left:* HIV prevalence: Blue: Data (with 95% confidence limits) and fitted line. Red: not on ART. Green: Data for those on ART and fitted line. *Middle:* HIV incidence. Red: Incidence of HIV. Green: incidence of treatment. Black: mortality not on ART. Grey: mortality on ART. *Right:* Cost of deaths: Red cost of letting people die without ART. Green cost of keeping people alive on ART. Blue: red plus green. Black: blue plus the cost of in-patient and out-patient care on and off ART[1] (see text for details). Brown: black plus the cost of testing.

While the current ART provision has saved many lives and greatly reduced incidence, there has also been a substantial cost saving; the annual cost to society has fallen from $50 per adult to $23 per adult (Figure 1F: black line) for a net cost saving of about $24M in the year 2013 alone and, even under the current policy of ART provision, costs will continue to fall.

Of the total cost in 2013 43% is the cost to society of people dying of AIDS related conditions (Figure 1F: red line), 14% is the cost of providing ART (Figure 1F: green line), and the remaining 43% is the cost of medical care (Figure 1F: black line).

**Universal Access to Early Treatment**

The substantial impact of treatment on the incidence, prevalence and mortality due to AIDS and the very substantial savings that accrue, encourages one to consider the impact of making ART available to everyone, regardless of CD4$^+$ cell counts. If the Gabonese government chooses to adopt the 2013 guidelines of the World Health Organization[7] about 90% of all HIV-positive people will be eligible for treatment as soon as they are found to be HIV-positive. The guidelines of the International AIDS Society[8] and the Department of Health and Human Services (DHHS)[9] both recommend treatment for those infected with HIV without regard to their CD4$^+$ cell count on the grounds that this is in the best interests of the individual concerned and has the added benefit of reducing the likelihood that they will not infect their partners.

We therefore consider what would happen under a policy of universal access in which 90% of all adults, not on ART, are tested each year and 90% of those that test positive are started on ART, with coverage and testing reaching half the target levels by the end of 2014.

Universal access to early treatment should eliminate HIV transmission (Figure 1G and H: red line) and end AIDS related deaths by 2022 (Figure 1H: black line) but there would still be a very large number of HIV positive people on ART (Figure 1G: blue line) who would have to be maintained on treatment for the rest of their lives. The rate at which people would be started on treatment will have to increase to about five times the present rate (Figure 1H: green line) in 2015 but after that would fall rapidly as transmission and the generation of new cases fell. Between 2015 and 2020 the cost would be only marginally greater than continuing the present policy (Figure 1F: black line *c.f.* Figure 1I: black and brown lines). After 2020 there would be further cost savings.

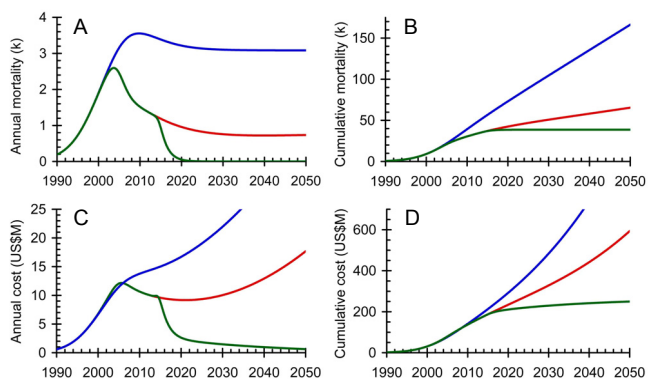

Figure 2. A: Total cost of ART and deaths discounted at 3.5% *p.a.* B: Annual adult AIDS deaths *per capita*. C: Cost per death averted. Blue line: null hypothesis if ART had not been made available; Red line: assuming current level of ART provision continued as in Figure 1D; Green line: early treatment as described in the text.

In Figure 2 we make a direct comparison of what would have happened without ART, what has happened as a result of the roll-out of ART, and what could happen under universal access and early treatment following the new WHO,[7] IAS[8] and DHHS[9] guidelines. The current policy has saved, and will continue to save, both money (Figure 2A) and lives (Figure 2B) in a cost effective manner (Figure 2C). Universal access to early treatment will cost marginally more between 2013 and 2015 but will cost less after that, will save many more lives and will lead to the elimination of HIV and the end of AIDS related deaths at a cost of less than $10 thousand per death averted and falling sharply as the epidemic is eliminated.

It is important to note that by 2020 testing costs alone would exceed all other costs (Figure 1I). It will be important to find more cost-effective ways of finding people who are HIV-positive either through expanded provider initiated counselling and testing or through contact tracing.

## Discussion

Gabon has been quite successful in the extent to which it has made ART therapy available in the public sector. Currently, an estimated 12 thousand people are being kept alive on ART with significant cost savings to the economy have been considerable. Expanding access will save more lives, will save more money and will lead to the elimination of HIV.

Although universal access to early treatment is the only way in which HV can be eliminated, other methods of support and control will play an important role.[3] The two most critical issues are drug supply and compliance. A regular and reliable supply of drugs must be assured. Stock-outs will create anger and mistrust among infected people and poor compliance, for any reason, will lead to viral rebound, treatment failure, ongoing transmission and drug resistance. These two considerations must be at the forefront of plans to effectively control and eventually eliminate HIV.

To ensure high levels of compliance it will be necessary to deal with problems of stigma and discrimination and to ensure that there is strong community involvement and support for people living with HIV. Adult medical male circumcision will significantly reduce the prevalence of HIV among young men and have a secondary benefit for young women who are the group at greatest risk. Pre-exposure prophylaxis will provide additional protection for those that are unable to protect themselves as is often the case for women who believe that their partners may be infected but are unable to negotiate condom use. Condoms are highly effective if used properly and should be readily available to all that need them. Better control of other sexually transmitted infections is important in itself and will contribute further to the control of HIV. And while these interventions can make an important contribution to stopping the epidemic of HIV, universal access to early treatment provides an ideal entry point for each of these interventions. Finally, by developing programmes that are firmly based in local communities it will be possible to provide training and education as well as jobs for community outreach workers thereby creating jobs and stimulating local economies.

It remains important to validate these assumptions, to ensure the validity or otherwise of these results and to improve the model predictions. To do this it is imperative that at the next ante-natal clinic survey, in October 2013, all women who test positive for HIV should also be tested for the presence of anti-retroviral drugs, their viral loads should be measured and an incidence assay should be used to estimate incidence,

What will be needed is a big push, hopefully when the Department of Health adopts the new WHO guidelines. Since an estimated 90% of all HIV positive people will then be eligible for immediate treatment it would be advisable to abandon the use of $CD4^{+}$ cell counts but to make viral load testing more widely available as a way of assessing the impact of the programme and of monitoring compliance. This analysis suggests that the rate at which people are currently being started on ART will have to increase by about four times, but only for about two years. While this will require considerable organization and a significant investment, the cost to the economy as a whole will be only marginally more than the cost of maintaining the present level of ART provision and within a few years the savings in both lives and money will be substantial.

We have the wherewithal to stop the epidemic all that is needed is the commitment to achieve an AIDS Free Generation.